\documentclass[twocolumn]{revtex4}
\usepackage{graphicx}
\usepackage{natbib}

\begin{document}

\title{$^2$H and $^{13}$C NMR studies on the temperature-dependent water and protein dynamics in hydrated elastin, myoglobin and collagen}

\author{Sorin A. Lusceac}
\affiliation{Institut f\"ur Festk\"orperphysik, Technische Universit\"at Darmstadt, Hochschulstr.\ 6, 64289 Darmstadt, Germany}
\author{Claudia R. Herbers}
\affiliation{Institut f\"ur Physikalische Chemie, Westf\"alische Wilhelms-Universit\"at M\"unster, Corrensstr.\ 28/30, 48149 M\"unster, Germany}
\altaffiliation[Present address: ]{Max Planck Institute for Polymer Research, Ackermannweg 10, 55128 Mainz, Germany}
\author{Michael R. Vogel}
\affiliation{Institut f\"ur Festk\"orperphysik, Technische Universit\"at Darmstadt, Hochschulstr.\ 6, 64289 Darmstadt, Germany}

\date{\today}

\begin{abstract}
$^2$H NMR spin-lattice relaxation and line-shape analyses are performed to study the temperature-dependent dynamics of water in the hydration shells of myoglobin, elastin, and collagen. The results show that the dynamical behaviors of the hydration waters are similar for these proteins when using comparable hydration levels of $h=0.25-0.43$. Since water dynamics is characterized by strongly nonexponential correlation functions, we use a Cole-Cole spectral density for spin-lattice relaxation analysis, leading to correlation times, which are in nice agreement with results for the main dielectric relaxation process observed for various proteins in the literature. The temperature dependence can roughly be described by an Arrhenius law, with the possibility of a weak crossover in the vicinity of $220\,$K. Near ambient temperatures, the results substantially depend on the exact shape of the spectral density so that deviations from an Arrhenius behavior cannot be excluded in the high-temperature regime. However, for the studied proteins, the data give no evidence for the existence of a sharp fragile-to-strong transition reported for lysozyme at about $220\,$K. Line-shape analysis reveals that the mechanism for the rotational motion of hydration waters changes in the vicinity of $220\,$K. For myoglobin, we observe an isotropic motion at high temperatures and an anisotropic large-amplitude motion at low temperatures. Both mechanisms coexist in the vicinity of $220\,$K. $^{13}$C CP MAS spectra show that hydration results in enhanced elastin dynamics at ambient temperatures, where the enhancement varies among different amino acids. Upon cooling, the enhanced mobility decreases. Comparison of $^2$H and $^{13}$C NMR data reveals that the observed protein dynamics is slower than the water dynamics.

\end{abstract}

\maketitle

\section{\label{sec:Intro}Introduction}

The biological function of many proteins depends on the existence of a hydration shell \cite{Rupley_APC_91}. Despite considerable progress in recent years, a complete understanding of the underlying coupling of protein and water motions is still lacking. In order to obtain fundamental insights, it proved useful to study the temperature dependence of protein and water dynamics, in particular through the protein dynamic transition at about $\mathrm{200\,K}$ \cite{Doster_EBPJ_08,Bagchi_CR_05}, which was found to be related to a freezing of biological functions \cite{Rasmussen_NAT_92}. The protein dynamics were proposed to be "slaved" \cite{Fenimore_PNAS_04} or "plasticized" \cite{Doster_EBPJ_08} by water dynamics, but the nature of the interplay of protein and water dynamics is still a subject of controversial debate.

Traditionally, protein dynamics were characterized on the basis of temperature-dependent atomic mean square displacements obtained from neutron scattering (NS) or M\"ossbauer experiments \cite{Doster_EBPJ_08}. Then, a crossover from linear behavior to non-linear behavior indicates an onset of non-harmonic motion upon heating, which was taken as evidence for the protein dynamic transition. More recently, combined NS and dielectric spectroscopy (DS) approaches were used to determine the temperature dependence of correlation times $\tau$ for protein dynamics \cite{Jansson_JNCS_06,Khodadadi_JPCB_08}. Since a discontinuity was not observed it was argued that a true dynamic transition does not exist, but the phenomena result from the protein dynamics entering the experimental time windows when the temperature is increased \cite{Khodadadi_JCP_08,Doster_EBPJ_08}.

Various experimental methods were employed to investigate the temperature-dependent dynamics of protein hydration waters. NS studies found a sharp crossover from a non-Arrhenius (fragile) behavior to an Arrhenius (strong) behavior upon cooling and attributed it to a fragile-to-strong transition (FST) at ca.\ $220\,$K \cite{Chen_PNAS_06}. Interestingly, the postulated FST occurs in the vicinity of the protein dynamical transition and a relation of both phenomena was postulated \cite{Chen_PNAS_06,Zanotti_EPJ_07}, in harmony with results from molecular dynamics (MD) simulations \cite{Kumar_PRL_06}. However, other studies, including DS work, did not observe a sharp transition, but a weaker crossover at lower temperatures or no crossover at all \cite{Swenson_PRL_04,Swenson_PRL_06,Sokolov_PRL_08,Cerveny_PRE_08,Ngai_JPCB_08}. These authors concluded that water dynamics, at least at sufficiently low temperatures, is not dominated by the structural ($\alpha$) relaxation, but a possibly largely universal \cite{Cerveny_PRE_08,Swenson_PRL_04} secondary ($\beta$) relaxation. For example, water dynamics was related to the Johari-Goldstein (JG) $\beta$ relaxation \cite{Ngai_JPCB_08}.

For a fundamental understanding of the interplay of protein and water dynamics, it is necessary, first, to assign the observed relaxation processes to the components and, second, to identify them as $\alpha$ or $\beta$ relaxations, requiring determination of the mechanisms for the respective motion. NMR experiments and MD simulations proved very useful tools for this purpose. In recent studies \cite{Vogel08,Vogel09}, we exploited the capabilities of these techniques to determine the temperature-dependent mechanisms for the dynamics of water in the hydration shells of elastin and collagen, two main components of the connective tissue. It was found that the mechanism for the water motion continuously changes from an isotropic motion to an anisotropic motion upon cooling. Moreover, the latter low-temperature process was shown to involve large-angle jumps. Such anisotropic large-angle jump motion is at variance with the mechanisms for the $\alpha$ and JG-$\beta$ relaxations of molecular supercooled liquids, suggesting the existence of a water-specific $\beta$ process. However, exact origin and possible universality of this $\beta$ process are open questions.

Here, we perform temperature-dependent $^2$H and $^{13}$C NMR experiments on hydrated proteins to investigate the water and the protein dynamics, respectively. On the one hand, we analyze $^2$H NMR spin-lattice relaxation (SLR) and $^2$H NMR line shape (LS) to study the respective water dynamics in the hydration shells of myoglobin, elastin, and collagen. In this way, we intend to address the question of universality. Moreover, for myoglobin, we perform a detailed comparison of the present results with literature data \cite{Swenson_data} to ascertain whether NMR and DS yield a coherent picture of water dynamics. On the other hand, protein dynamics is investigated in $^{13}$C NMR using cross polarization (CP) and magic-angle spinning (MAS) techniques. Specifically, we record $^{13}$C CP MAS spectra of elastin to monitor site-resolved changes of the protein dynamics upon hydration and cooling, respectively.

\section{\label{sec:Mat}Materials and methods}

\subsection{\label{subsec:Mat} Materials and experiments}

We study collagen, elastin, and myoglobin hydrated to different levels $h$ (g water per 1 g protein) using D$_2$O ($^2$H NMR) or H$_2$O ($^{13}$C NMR). The samples are referred to as C (collagen), E (elastin) or M (myoglobin) followed by the hydration level in percent. Details about the collagen and elastin samples can be found in Ref.\ \cite{Vogel08}. M35 was prepared from sperm-whale myoglobin (Sigma) and kindly provided by W.\ Doster.

$^2$H NMR SLR and LS measurements were performed at Larmor frequencies $\omega_0$ of $2\pi\times 76.8$\,MHz for collagen and elastin samples and $2\pi\times 46.1$\,MHz for M35. The former setup is described in Ref.\ \cite{Vogel06}. In the latter setup, a home-built probe was placed in a Konti CryoVac cryostat operating with liquid nitrogen and controlled by a TIC 304 MA CryoVac temperature controller. The temperature stability was better than $0.5$\,K. A $90^{\circ}$ pulse length of $2.1$\,${\mu}$s was used, ensuring adequate excitation of the broad spectrum. The solid-echo pulse sequence, $90^\circ_x$ -- $t_p$ -- $90^\circ_{y}$, preceded by a saturation-pulse sequence, was employed to acquire $^2$H NMR spectra. The $^2$H NMR SLR was measured utilizing the saturation-recovery technique in combination with a solid echo for detection. The $^{13}$C CP MAS NMR spectra were recorded with a Bruker DSX 400 spectrometer working at $\omega_0=2\pi\times 100.6$\,MHz and a 4-mm triple-resonance Bruker MAS probe. For CP, a $^1$H $90^\circ$ pulse of $3.5\,\mu$s length was followed by a contact time of $1\,$ms. Two-pulse phase modulation (TPPM) decoupling \cite{TPPM} with a field strength of $71\,$kHz was applied. The spinning speed was $8$\,kHz. The $^{13}$C chemical shifts were referenced with respect to tetramethylsilane. A flow of nitrogen gas controlled by a Bruker VT 3000 heating unit was used to adjust sample temperature. Since set and actual temperature differ substantially in MAS experiments, deviations were removed within an uncertainty of $\pm1$\,K by means of temperature calibration with $^{207}$Pb NMR of lead nitrate \cite{Pb}.

\subsection{\label{subsec:Theo} Theoretical background}

In solid-state $^2$H NMR, the dominant interaction is the first-order quadrupolar interaction, which describes the coupling between the electric quadrupole moment of the deuteron and the electric field gradient (EFG) at the nuclear site. The corresponding quadrupolar contribution to the NMR frequency is given by \cite{Spiess}:
\begin{equation}
  \label{eq:Theo_omega}
  \omega_Q(\theta,\phi)=\pm\frac{\delta}{2}\,
  [3\cos^2(\theta)-1-\eta\sin^2(\theta)\cos(2\phi)].
\end{equation}
Here, $\theta$ and $\phi$ are the polar coordinates of the external magnetic flux density $\textbf{B}_0$ in the principal axis system of the EFG tensor, $\delta$ and $\eta$ are the anisotropy parameter and the asymmetry parameter, respectively, and the "$\pm$" sign corresponds to the two allowed transitions of the $I=1$ nucleus. In our case of D$_2$O, the asymmetry of the EFG tensor is negligibly small ($\eta\approx0.1$), leading to
\begin{equation}
  \label{eq:Theo_omega_simple}
  \omega_Q(\theta)=\pm\frac{\delta}{2}\,
  [3\cos^2(\theta)-1],
\end{equation}
where $\theta$ is the angle between $\textbf{B}_0$ and the direction of the O--D bond. Thus, the orientation of the water molecule determines the resonance frequency, rendering $^2$H NMR a precious tool to study water rotational dynamics.

If the correlation time of water reorientation exceeds the inverse anisotropy parameter, $\tau\gg1/\delta$, the anisotropy of the quadrupolar interaction and the powder average will result in broad $^2$H NMR spectra for disordered materials \cite{Pake48}. When the temperature is increased and, hence, molecular dynamics becomes faster, these static spectra, which are called Pake spectra, collapse as soon as $\tau$ becomes comparable to $1/\delta$, i.e., for molecular dynamics in the microseconds regime. Finally, motionally averaged spectra are observed in the limit of fast motion, $\tau\ll1/\delta$. In particular, fast isotropic reorientation results in narrow Lorentzian lines. When a broad distribution of correlation times $G(\log \tau)$ exists so that there are both slow ($\tau\gg1/\delta$) and fast ($\tau\ll1/\delta$) molecules at a given temperature, the LS will be comprised of a Pake spectrum and a motionally narrowed line \cite{Roessler90}. Such spectra have been called two-phase spectra.

For isotropic motions, the $^2$H SLR time $T_1$ is given by \cite{Boehmer01,Bloembergen48}
\begin{equation}
  \label{eq:BPP}
  \frac{1}{T_1}=\frac{2}{15}\,{\delta}^2\left[J(\omega_0)+4J(2\omega_0)\right],
\end{equation}
where $J(\omega)$ is the spectral density of the molecular reorientation. Hence, SLR analysis provides access to the time scale of molecular rotational motion, e.g., $T_1$ is a minimum when $\tau\approx 1/\omega_0$, i.e., for molecular dynamics in the nanoseconds range.

Quantitative SLR analysis requires knowledge about the shape of the corresponding rotational autocorrelation function $F_2(t)$. For a Debye process, $F_2(t)=\exp(-t/\tau)$. Then, the spectral density can be written as
\begin{equation}
  \label{eq:sp_dens_Debye}
  J_{BPP}(\omega)=\frac{\tau}
  {1+(\omega\tau)^2}
\end{equation}
and Eq.\ (\ref{eq:BPP}) takes the form derived by Bloembergen, Purcell, and Pound (BPP) \cite{Bloembergen48}. For disordered materials, existence of distributions $G(\log \tau)$ usually leads to a nonexponentiality of $F_2(t)$. In the literature \cite{Boehmer01,Beckmann88}, different spectral densities were used to consider such distributions. For various hydrated proteins, DS studies found that the loss peak, which was ascribed to water dynamics, is symmetric and well described by the empirical Cole-Cole (CC) distribution function \cite{Swenson_PRL_06,Swenson_data,Cerveny_PRE_08,Swenson_PRL_04,Khodadadi_JPCB_08}. Therefore, we use the CC spectral density \cite{Beckmann88}

\begin{equation}
  \label{eq:sp_dens_CC}
  J_{CC}(\omega)=\frac{\sin(\frac{\pi}{2}\alpha_{cc})(\omega\tau_{cc})^{\alpha_{cc}}}{\omega[1\!+\!(\omega\tau_{cc})^{2\alpha_{cc}}\!+\!2\cos(\frac{\pi}{2}\alpha_{cc})(\omega\alpha_{cc})^{\alpha_{cc}}]}
\end{equation}
in our $^2$H SLR analysis. Here, $\tau_{cc}$ is the position of the maximum of $G(\log \tau)$ and $\alpha_{cc}$ describes the width of this symmetric distribution. Near ambient temperatures, information about the shape of the spectral density from DS is lacking. In this temperature range, we also employ the Cole-Davidson (CD) spectral density
\begin{equation}
  \label{eq:sp_dens_CD}
  J_{CD}(\omega)=\frac{\sin[\beta_{cd}\,{\arctan}(\omega\tau_{cd})]}
  {\omega[1+(\omega\tau_{cd})^2]^{\beta_{cd}/2}}.
\end{equation}
for $^2$H SLR analysis. The asymmetric CD distribution is characterized by a mean correlation time $\overline{\tau}_{cd} =\tau_{cd}\beta_{cd}$ \cite{Beckmann88}.    

\section{\label{sec:Res} Results}

\subsection{$^2$H NMR SLR}

\begin{figure}[tp]
  \centering
   \includegraphics[width=8cm,clip]{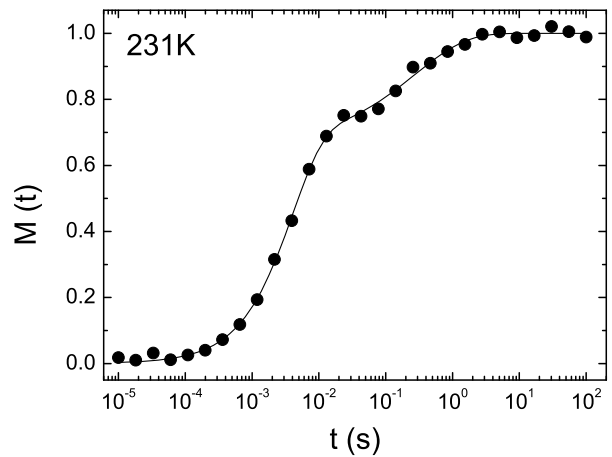}
\caption{\label{fig:magnetization}
Buildup of the normalized equilibrium magnetization after saturation for M35 at 231K. The line is a fit with Eq.~(\ref{eq:magnetization})}
\end{figure}

Figure~\ref{fig:magnetization} shows a typical buildup of the normalized equilibrium magnetization $M(t)$ after saturation for M35. We see a two-step shape, which is a consequence of deuteron-proton exchange between water and myoglobin molecules \cite{Ellis76}: exchangeable hydrogens of the protein are partially replaced by deuterons so that the $^2$H NMR signals are comprised of the respective contributions from water deuterons and myoglobin deuterons. The observation of distinct steps means that the deuteron-proton exchange is slow on the time scale of the buildup and, hence, we can separately study the dynamical behaviors of the components in $^2$H NMR. In previous work on samples with higher hydration levels of $h\geq0.6$ \cite{Vogel08}, we observed a third step at long times, which was attributed to crystalline water. For M35, we find no evidence for the existence of such third step and, thus, of crystalline water.

Considering the two-step signature of the buildup, we fit the data to
\begin{equation}
  \label{eq:magnetization}
M(t)=1-\left[a_we^{-\left(\frac{t}{T_{1w}}\right)^{\beta_w}}+(1-a_w)e^{-\left(\frac{t}{T_{1p}}\right)^{\beta_p}}\right].
\end{equation}
Here, $T_{1w}$ and $T_{1p}$ are the SLR times of water and protein, respectively, and $a_w$ and $(1-a_w)$ describe the heights of the steps. To take into account that $^2$H NMR SLR relaxation can be nonexponential for disordered materials \cite{Boehmer01}, we introduce the stretching parameters $\beta_w$ and $\beta_p$ and calculate the mean SLR times as
\begin{equation}
  \label{eq:meanT1}
\langle T_{1w,1p}\rangle =\frac{T_{1w,1p}}{\beta_{w,p}}\,\,\Gamma\left(\frac{1}{\beta_{w,p}}\right),
\end{equation}
where $\Gamma$ is the Gamma function. The indices $w$ and $p$ denote water and protein, respectively.

\begin{figure}[tp]
  \centering
   \includegraphics[width=8cm,clip]{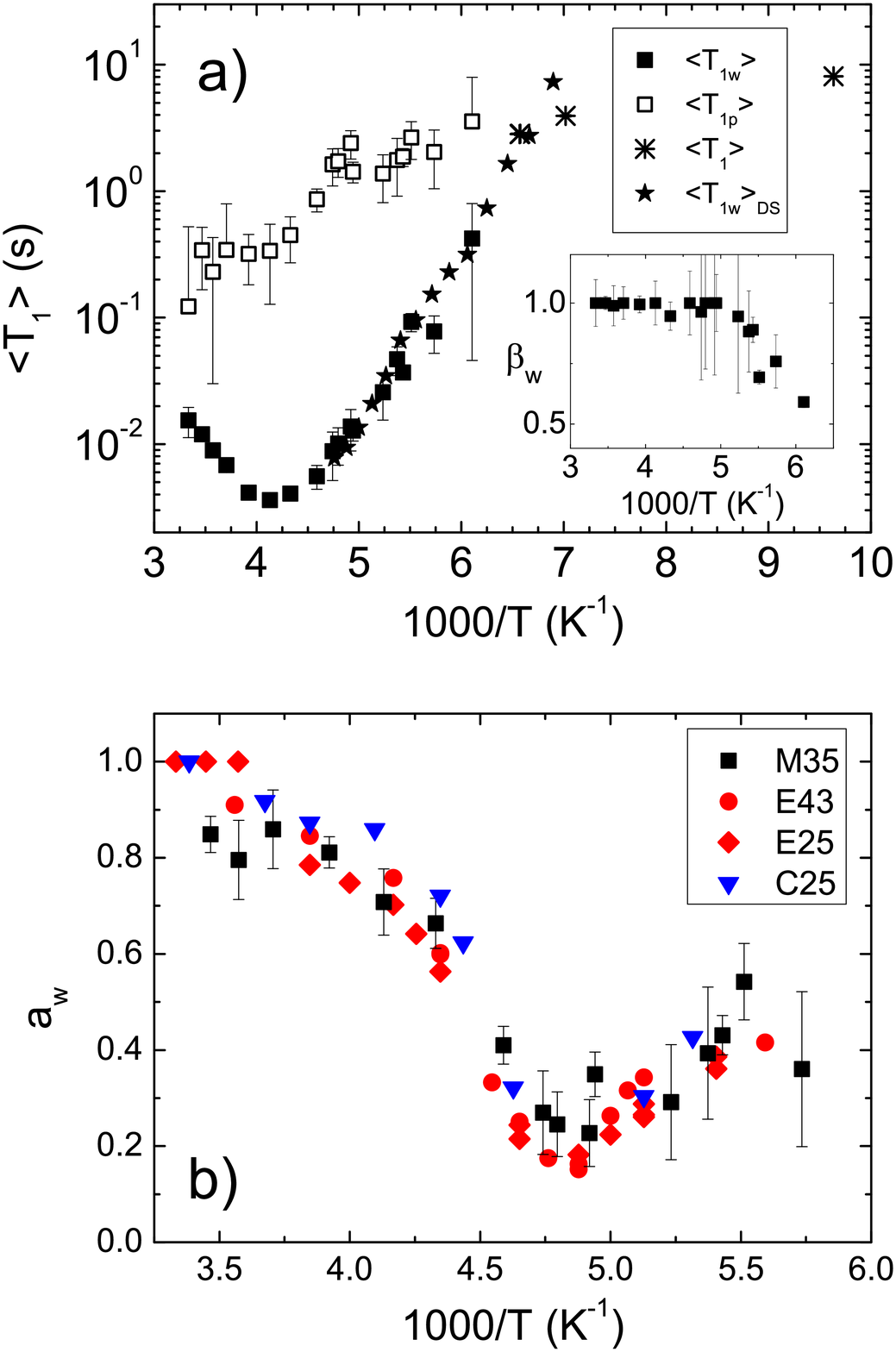}
\caption{\label{fig:T1andAW}
a) Arrhenius plot of the mean SLR times $\langle T_{1w} \rangle$ and $\langle T_{1p} \rangle$ of water and protein deuterons in M35, respectively. $\langle T_{1} \rangle $ is the mean SLR time in the temperature range $T<155\,$K where one-step magnetization buildup is observed. The experimental results are compared with SLR times $\langle T_{1w}\rangle_{DS}$, which were calculated using the distributions $G(\log \tau)$ from DS on M45 \cite{Swenson_data}. Inset: stretching parameter $\beta_w$ characterizing the nonexponentiality of the SLR step from water deuterons. b) Temperature dependence of the relative heights $a_w$ of the water signals for M35, E25, E43, and C25.}
\end{figure}

The results of this analysis enable unambiguous assignment of the steps to water and protein, see Fig.\ \ref{fig:T1andAW}. The SLR time of one component is short and and exhibits a pronounced minimum, whereas that of the other component is long and increases upon cooling in the whole studied temperature range. While the latter behavior indicates a slow ($\tau\gg1/\omega_0$) component and, hence, the long-time step can be assigned to the protein, the former behavior is characteristic of a highly mobile component so that the short-time step results from the water. This assignment is confirmed by the relative heights of the relaxation steps at ambient temperatures. On the basis of the amino acid composition of myoglobin \cite{Doster_BBA_05}, it is expected that water deuterons substantially outnumber myoglobin deuterons for M35. Therefore, the higher first and lower second steps result from water and protein, respectively. The pronounced temperature dependence of $a_w$ is an NMR effect, as will be discussed below.

Having assigned the steps, the SLR of water and myoglobin can be discussed in more detail. For water, we find a minimum of $\langle T_{1w} \rangle$ at about $240\,$K, indicating water dynamics with $\tau\approx 1/\omega_0$. Important information about a possible existence of a distribution $G(\log \tau)$ is offered by the value at the minimum, $\langle T_{1w}\rangle_{m}$. It is straightforward to calculate $\langle T_{1w}\rangle_{m}$ for a Debye process, see Eq.\ (\ref{eq:sp_dens_Debye}). Using $\omega_0=2\pi\times 46.1$\,MHz and $\delta=2\pi\times166\,$kHz, as determined from the low-temperature spectrum in Fig.~\ref{spectraM}, a minimum value of $1.4\,$ms is expected, which is about a factor of 2 smaller than the experimental result $\langle T_{1w}\rangle_m=3.6\,$ms. This discrepancy means that a Debye process does not describe the water dynamics, but a distribution $G(\log \tau)$ exists \cite{Boehmer01}. For myoglobin, $\langle T_{1p} \rangle$ shows a weak increase with decreasing temperature, implying $\tau\gg 1/\omega_0$.  Below $160\,$K, equilibrium magnetization is build up in a single step and, thus, water and myoglobin exhibit a common SLR time of a few seconds. This means that an exchange of magnetization occurs between the components on this time scale. Possible candidates for exchange processes are spin diffusion, i.e., a transfer of magnetization due to flip-flop processes of the spins, and deuteron exchange.

Next, we turn to the stretching parameters $\beta_w$ and $\beta_p$, describing deviations from exponential SLR. For water, we find exponential SLR above $200\,$K, while cooling results in a decrease of the stretching parameter $\beta_w$ at lower temperatures, indicating increasingly nonexponential SLR, see Fig.\ \ref{fig:T1andAW}(a). In our case, nonexponential SLR is expected since the existence of a distribution of correlation times should lead to a distribution of SLR times, $V(T_1)$. Therefore, our findings imply that, above $200\,$K, water dynamics averages over any distribution $V(T_1)$, i.e., each water molecule samples a representative set of local environments on the time scale of $T_{1w}$, leading to exponential SLR. When the temperature is decreased, the time scale of water dynamics and, thus, of exchange processes within the distribution $G(\log \tau)$ approaches that of SLR so that the averaging process becomes incomplete and the existence of a distribution $V(T_1)$ is revealed, i.e., SLR becomes nonexponential. Such temperature dependence of the stretching parameter is known from $^2$H NMR SLR studies on the $\alpha$ process of supercooled liquids \cite{Boehmer01}. For myoglobin, the stretching parameter $\beta_{p}=0.6$ is independent of temperature within the accuracy of the measurement. Thus, the protein dynamics are heterogeneous and the corresponding distribution $V(T_1)$ is not averaged due to exchange processes within the distribution $G(\log \tau)$.

Before we continue our SLR analysis, it is instructive to determine whether NMR and DS studies yield consistent results for the dynamics of myoglobin hydration water. For this purpose, we directly use the CC distributions obtained from a DS study on M45 \cite{Swenson_data} to calculate the SLR times according to Eq.\ (\ref{eq:sp_dens_CC}). The resulting values $\langle T_{1w}\rangle_{DS}$ are included in Fig.\ \ref{fig:T1andAW}(a). We see that the calculated data nicely agree with the NMR results in the common temperature range below the $\langle T_{1w} \rangle$ minimum. Thus, NMR and DS probe the same dynamical process, which is appropriately described by a CC distribution, at least below $220\,$K.

\begin{figure}
  \centering
   \includegraphics[width=8cm,clip]{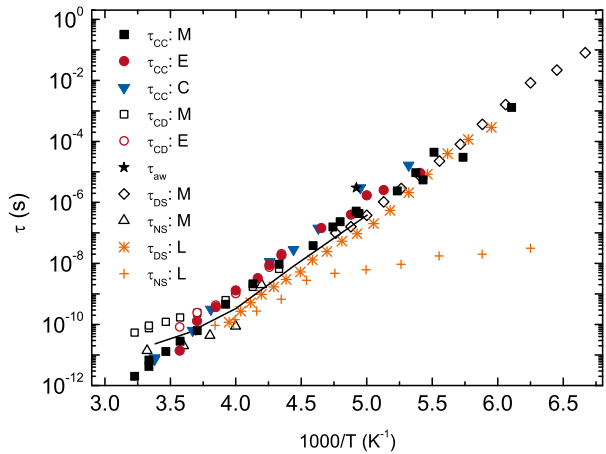}
\caption{\label{corrtimes}
Correlation times of protein hydration waters. The values of $\tau_{cc}$, $\overline{\tau}_{cd}$, and $\tau_{aw}$ from the present SLR analyses for water in the hydration shells of myoglobin (M35), elastin (E25), and collagen (C25) are compared with literature data for myoglobin and lysozym (L) hydration waters. Specifically, correlation times from DS on M45 \cite{Swenson_data}, NS on M35 \cite{Doster_BBA_05}, DS on L37 \cite{Khodadadi_JPCB_08}, and NS on L30 \cite{Chen_PNAS_06} are shown. For comparison, correlation times of protein dynamics from NS on L40 \cite{Khodadadi_JPCB_08} are included as line.}
\end{figure}

For further SLR analysis, we first assume that a CC distribution describes water dynamics at all studied temperatures. In DS \cite{Swenson_data}, an essentially temperature independent width parameter $\alpha_{cc}\approx0.48$ was found for M45. In $^2$H NMR, the width parameter is determined by the value of $\langle T_{1w}\rangle_{m}$. We obtain $\alpha_{cc}=0.45$ at $240\,$K, in harmony with the result from DS. Assuming that $\alpha_{cc}$ does not depend on temperature, we can extract $\tau_{cc}$ from $\langle T_{1w} \rangle$. In Fig.~\ref{corrtimes}, we compare the temperature dependence of $\tau_{cc}$ for hydrated myoglobin, elastin and collagen. The values for the two latter proteins are obtained when we use the CC spectral density ($\alpha_{cc}=0.50$) to analyze the SLR times presented in previous work \cite{Vogel08}. We see that the hydration waters of all studied proteins exhibit similar correlation times, which roughly follow an Arrhenius law. Closer inspection of the data implies that there may be a weak change of the temperature dependence at about $220\,$K. Consistently, separate fits of the myoglobin data in the ranges $T>220\,$K and $T<220\,$K yield activation energies of $E_a=0.66\,$eV and $E_a=0.55\,$eV, respectively.

Since information from DS about the shape of the spectral density is lacking for myoglobin above $220\,$K, we determine to which extent SLR analysis depends on the choice of $J(\omega)$ in this temperature range. Using $J_{CD}(\omega)$, a width parameter $\beta_{cd}=0.19$ results from the value of $\langle T_{1w}\rangle_{m}$. Assuming a temperature independent width of the CD distribution, we determine the mean correlation times $\overline{\tau}_{cd}$ from $T_{1w}$. The results are included in Fig.\ \ref{corrtimes}. We see that the CC and CD spectral densities yield comparable correlation times in the range $230-270\,$K, i.e., in the vicinity of the $\langle T_{1w} \rangle$ minimum, while the findings differ substantially near ambient temperatures. In particular, non-Arrhenius behavior is observed for $\overline{\tau}_{cd}$.  

Figure \ref{fig:T1andAW}(b) shows the temperature-dependent relative height $a_w$ of the water signal in M35 together with data for E25, E43 and C25. For the studied samples, $a_w$ exhibits a comparable temperature dependence with a minimum at about $205\,$K. This minimum can be explained by the use of a solid echo for signal detection in our experiments, see Sec.\ \ref{subsec:Mat}. It indicates that water dynamics during the dephasing and rephasing periods $t_p=20\,\mu$s of the solid-echo sequence interferes with echo formation \cite{Spiess81}. This conclusion is confirmed by the results of our SLR analysis, which shows that water dynamics is indeed characterized by $\tau$ of the order of a few microseconds at the minimum position of $205\,$K, see Fig.\ \ref{corrtimes}. By contrast, the signal is not reduced for much faster and slower dynamics when $\omega_Q$ is time-averaged and time-invariant, respectively. Detailed random-walk simulations for different motional models show that the solid-echo amplitude is a minimum for a correlation time of about $3\,\mu$s \cite{Vogel00,Vogel09}. Hence, the coincidence of the minima of $a_w$ shows that the hydration waters of myoglobin, elastin and collagen exhibit a correlation time $\tau_{aw}=3\,\mu$s at $205\,$K. In Fig.\ \ref{corrtimes}, we see that $\tau_{aw}$ is comparable to $\tau_{cc}$, indicating that both methods yield consistent results.

\subsection{$^2$H NMR line-shape analysis}
\begin{figure}
  \centering
   \includegraphics[width=8cm,clip]{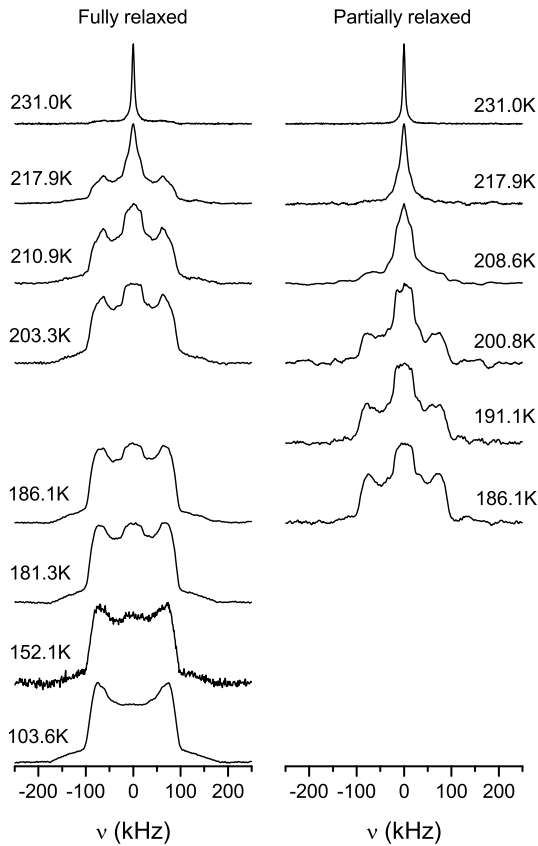}
\caption{\label{spectraM}
Solid-echo spectra of M35 at various temperatures. Left column: fully relaxed spectra comprising signals  from myoglobin and water deuterons. Right column: partially relaxed spectra to single out the signal from water deuterons, see text for details.}
\end{figure}

In order to gain information about the mechanisms for water dynamics, we analyze the temperature-dependent $^2$H NMR LS. Results for M35 are presented in Fig.~\ref{spectraM}. Fully relaxed (FR) and partially relaxed (PR) spectra are shown in the left and right columns, respectively. When measuring FR spectra, we wait for complete recovery of magnetization after saturation, while the waiting time between the saturation and solid-echo sequences was set to $\langle T_{1w} \rangle$ when acquiring PR spectra. Thus, both myoglobin and water contribute to the FR spectra, while the water component is singled out in the PR spectra. Comparison of the FR and PR spectra shows that, at all studied temperatures, myoglobin adds a broad Pake spectrum to the LS, indicating that the deuterated regions of the protein show rotational motion that is slow ($\tau\gg1/\delta$) or has small angular amplitude ($<15^\circ$). Moreover, we see that the relative contribution of myoglobin to the FR spectra is particularly prominent in the vicinity of $205\,$K, where the contribution of water is reduced because of dynamics during the solid-echo sequence, see Fig.\ \ref{fig:T1andAW}(b).

Focusing on the PR spectra, we see that, below $200\,$K, the spectral contribution of water is comprised of a broad Pake spectrum and a central feature, the prominent intensity of which extends ca.\ from $-20$ to $20\,$kHz. With decreasing temperature, the intensity of the latter component slowly decreases, while its shape does not change. Unfortunately, it is not possible to single out the water signal in PR spectra below $186\,$K because the difference between $T_{1w}$ and $T_{1p}$ diminishes. However, it is evident from the FR spectra at low temperatures, that the central feature continues to loose intensity upon further cooling until it disappears between 100 and $150\,$K. The observation that, below $200\,$K, the central part of the spectrum changes intensity, but not shape shows that it is a fast-motion limit spectrum, i.e., it results from water molecules exhibiting $\tau\ll1/\delta$. The shape of this component clearly deviates from a Lorentzian, indicating that the underlying motion is anisotropic. The coexistence of spectra typical of the limits of fast and slow motion implies that a broad distribution $G(\log \tau)$ governs the anisotropic water reorientation at $T<200\,$K. Then, the slow ($\tau\gg1/\delta$) water molecules from the distribution give rise to the Pake spectrum, while the fast ($\tau\ll1/\delta$) ones yield the central feature between $-20$ and $20\,$kHz. When temperature is varied and, hence, $G(\log \tau)$ shifts, the relative intensity of both spectral components continuously changes, while the respective LS does not \cite{Roessler90}.

Above $205\,$K, a narrow Lorentzian line emerges, indicating that a fraction of water molecules start exhibiting fast ($\tau\ll1/\delta$) isotropic motion, or at least fast motion with cubic symmetry. Interestingly, both fast-motion limit spectra, the Lorentzian and the central feature between $-20$ and $20\,$kHz coexist in the temperature range $208-218$\,K. Hence, there are water molecules that exhibit fast isotropic motion while others show fast anisotropic motion, but no fast isotropic motion. This means that two different mechanisms exist for the rotational motion of hydration water. Upon heating, the intensity of the Lorentzian rapidly increases until the PR spectra are well described by a single Lorentzian above $230\,$K. Hence, more and more water molecules exhibit fast isotropic motion and, eventually, all molecules are involved in such dynamics in the high-temperature range.

The shape of the central feature provides important insights into the mechanism for water rotational motion below $200\,$K. The main intensity of this LS component extends from $-20$ to $20\,$kHz and minor intensity is found up to about $\pm 40\,$kHz. Thus, as compared to the static case, the width of the spectrum is reduced by more than a factor of 4. Such pronounced reduction of the spectral width cannot result from small-amplitude motion, but it rather requires large-amplitude motion. Furthermore, two-site jumps lead at most to a motional narrowing by a factor of 2 \cite{Spiess}. Hence, the small width of the central feature provides strong evidence against the possibility that the water molecules perform exact $\pi$ flips about their $C_2$ symmetry axes. For fast uniaxial rotation or for fast jumps between $N$ sites ($N>3$) with $C_N$ symmetry, the motionally averaged quadrupolar coupling tensor is characterized by an anisotropy parameter $\overline{\eta}=0$. Moreover, since $\eta\approx 0$ in the static case, the anisotropy parameter $\overline{\delta}$ of the averaged tensor is given by \cite{Spiess}
\begin{equation}
\frac{\overline{\delta}}{\delta}=\frac{1}{2}\left(3\cos^2\beta-1\right).
\end{equation}
Here, $\beta$ is the angle between the rotation axis and the possible orientations of the O--D bonds. Then, our case of $|\overline{\delta}|\approx 2\pi\times40\,$kHz translates into $\beta\approx45^\circ$ or $\beta\approx65^\circ$. The ambiguity results because the symmetry of the LS hampers determination of the sign of the asymmetry parameter in $^2$H NMR. Thus, within these models, the maximum angular displacement of the O--D bonds, which corresponds to the full opening angle $2\beta$ of the cone, amounts to $90^\circ$ or $130^\circ$. 

However, the observed central feature cannot be described by a single Pake spectrum, but it is comprised of a distribution of spectral shapes, corresponding to a distribution of geometries of water motion, which may be expected for a disordered material. For example, a fraction of the motionally averaged tensors may exhibit asymmetry parameters $\overline{\eta}>0$. Therefore, the present results are at variance with exact $\pi$ flips, but not with strongly distorted $\pi$ flips. Also, LS analysis allows us to rule out exact tetrahedral jumps, which would lead to a narrow Lorentzian in the limit of fast motion, but strongly distorted tetrahedral jumps cannot be excluded. Altogether, the shape of the central feature does not enable determination of a unique motional model, but it nevertheless indicates that, below $200\,$K, myoglobin hydration water performs anisotropic reorientation, which covers large angular displacements, but deviates from exact two-site or tetrahedral jumps.

\begin{figure}
  \centering
   \includegraphics[width=8cm,clip]{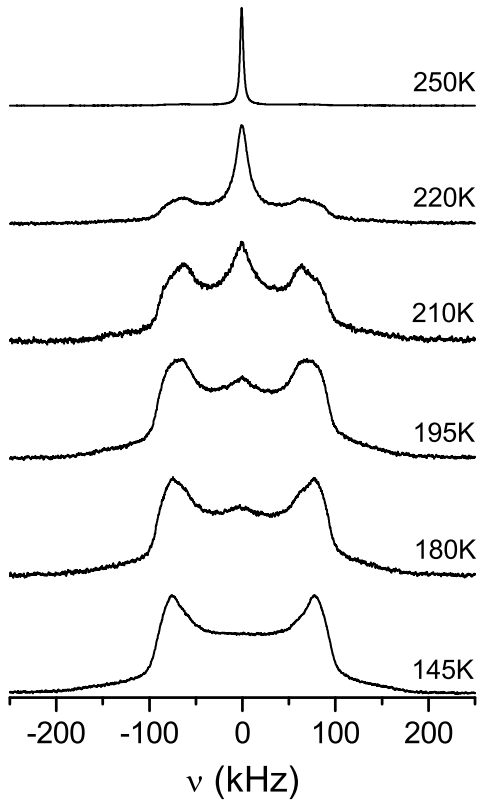}
\caption{\label{spectraE}
Solid-echo spectra of E25 at several temperatures (fully relaxed).}
\end{figure}

Finally, we study whether the $^2$H NMR LS differs for the hydration waters of various proteins. FR spectra of E25 are shown in Fig.~\ref{spectraE}. We see that the limit of slow motion, $\tau\gg1/\delta$, is found somewhat below $145\,$K. As the temperature is increased, a central, broad feature gains intensity and, finally, the LS is dominated by a narrow Lorentzian. Thus, the results for E25 resemble that for M35, see Fig.\ \ref{spectraM}. However, the shape of the central feature is different for both proteins. In particular, for E25 and E43 \cite{Vogel09} at $T= 210-220\,$K, the data give no evidence for a coexistence of spectral components indicative of fast isotropic and fast anisotropic motions, respectively. The origin of this difference is not yet understood. It may indicate that the mechanisms for low-temperature rotational motion differ between the hydration waters of elastin and myoglobin. Alternatively, a crossover from isotropic to anisotropic motion may occur at larger values of $\tau$ for elastin than for myoglobin so that the limit of fast motion cannot be reached for the anisotropic dynamics of elastin hydration waters.
\subsection{$^{13}$C CP MAS NMR spectra}

\begin{figure}
  \centering
   \includegraphics[width=8cm,clip]{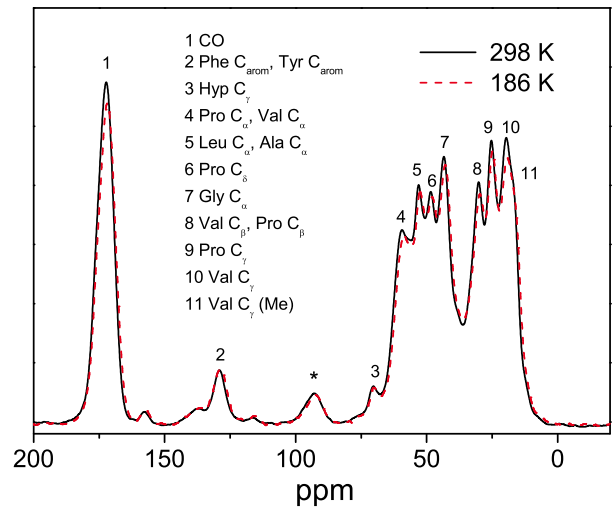}
\caption{\label{13C_E0}
$^{13}$C CP MAS NMR spectra of E0 at various temperatures. The assignment of the lines follows Ref. \cite{Kumashiro} }
\end{figure}

\begin{figure}
  \centering
   \includegraphics[width=8cm,clip]{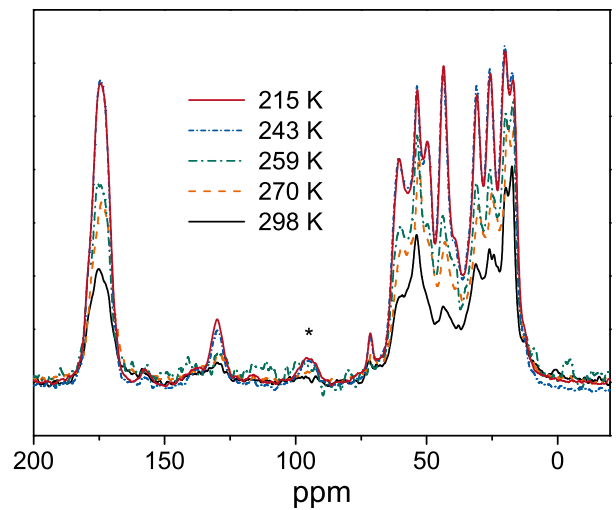}
\caption{\label{13C_E60}
$^{13}$C CP MAS NMR spectra of E60 at various temperatures.}
\end{figure}

To investigate the dependence of elastin dynamics on hydration and temperature, we use $^{13}$C NMR. Figures \ref{13C_E0} and \ref{13C_E60} show temperature-dependent $^{13}$C CP MAS spectra of E0 and E60, respectively. The data were divided by the number of used accumulations and corrected for the Curie factor, enabling straightforward comparison. For E0, the results are independent of temperature in the studied range 186 -- $298\,$K, indicating that substantial elastin dynamics is absent on the experimental time scale. By contrast, a strong temperature dependence of the spectra is evident for E60. At $298\,$K, the spectral intensity is significantly reduced and the relative intensities of the various peaks differ from that found for E0. In $^{13}$C CP MAS, such reduction of the peak intensities results when molecular dynamics in the sub-milliseconds regime reduce the strength of the $^1$H--$^{13}$C dipolar couplings, which are required for effective polarization transfer. Therefore, the LS of E60 shows an enhancement of elastin dynamics due to hydration near room temperature, consistent with findings in previous works \cite{Perry_BJ_02,Yao_MRC_04}. The strength of this effect differs for different types of carbons. It is prominent for Gly C$_\alpha$ and Pro $C_{\alpha,\beta,\gamma}$, implying that the dynamics of Gly and Pro amino acids are particularly sensitive to hydration. When the temperature is decreased, the LS of E60 continuously approaches the temperature-independent LS of E0 and, eventually, the spectra of both samples are very similar at $T\leq243\,$K. Thus, the enhanced elastin mobility of the wet sample diminishes. Consistently, the results of the present and previous \cite{Vogel08} $^2$H NMR studies on hydrated myoglobin, elastin and collagen revealed an increase of $\langle T_{1p} \rangle$ upon cooling, see Fig.\ \ref{fig:T1andAW}, implying a continuous reduction of protein mobility. 

\section{\label{sec:Dis} Discussion}

We have used $^2$H and $^{13}$C NMR to study temperature-dependent water and protein dynamics, respectively. $^2$H NMR LS and SLR analyses have revealed that, despite minor quantitative differences, water shows comparable rotational dynamics in the hydration shells of collagen, elastin, and myoglobin. For all studied materials, water reorientation cannot be described as a Debye process, but broad distributions of correlation times $G(\log \tau)$ exist. 

To consider the existence of such distributions, SLR analysis has mainly been performed using the CC spectral density. This choice has been motivated by the observation of symmetric dielectric loss peaks for various hydrated proteins \cite{Swenson_PRL_06,Cerveny_PRE_08,Khodadadi_JPCB_08}. The mean correlation times $\tau_{cc}$ and width parameters $\alpha_{cc}$ obtained from such SLR analyses are in good agreement for the hydration waters of collagen, elastin, and myoglobin. The temperature dependence of $\tau_{cc}$ is roughly described by an Arrhenius law, although there is some evidence for a weak crossover at about $220\,$K. The width parameters $\alpha_{cc}\approx0.48$ correspond to full widths at half maximum of about 3.3 orders of magnitude. Near ambient temperatures, DS does not provide information about the shape of the spectral density. Therefore, we also employed the CD spectral density for SLR analysis. While the correlation times from $J_{CD}$ and $J_{CC}$ are comparable in the range $230-270\,$K, they differ at higher temperatures. In particular, application of the CD spectral density leads to deviations from Arrhenius behavior in the high-temperature range.

LS analysis has revealed that the mechanism for rotational motion of protein hydration waters changes in the vicinity of $220\,$K, where the temperature-dependent correlation times may exhibit a weak crossover. The nature of these changes differ somewhat in the hydration shells of diverse proteins. For myoglobin hydration water, we have observed an isotropic motion at high temperatures and an anisotropic motion at low temperatures. Interestingly, both mechanisms coexist in the vicinity of $220\,$K, suggesting that two different dynamical processes exist. Moreover, LS analysis has shown that the anisotropic motion has large amplitude. Specifically, it involves angular displacements on the order of the tetrahedral angle. Our findings exclude that water performs exact two-site or tetrahedral jumps. However, distorted two-site jumps, including inexact $\pi$ flips, or distorted tetrahedral jumps may be consistent with the data, provided there is a substantial distribution of geometries of the motion. Although LS analysis does not enable determination of a unique model of low-temperature water reorientation, it allows us to draw some important conclusions. Because of its anisotropy, the observed motion is not related to the $\alpha$ process of myoglobin hydration water. Moreover, the large-amplitude of the reorientation is strong evidence against the conjecture \cite{Ngai_JPCB_08} that the low-temperature process can be identified with the JG-$\beta$ process, which results from small-amplitude motion for molecular glass formers \cite{Vogel01}. Therefore, we propose that a water-specific $\beta$ process is observed, consistent with the conclusions in our previous experimental and computational studies \cite{Vogel08,Vogel09}. Specifically, in MD simulation work, we found that the mechanism for water motion changes from small-angle (diffusive) motion to large-angle (jump) motion \cite{Vogel09}. In future work, we plan to use two-dimensional $^2$H NMR in time and frequency domains to further narrow down the possible mechanisms for the rotational dynamics of protein hydration waters at low temperatures.

It is instructive to compare the present results with literature data for the dynamics of
protein hydration waters. First, we emphasize that, although there may be a weak
crossover in the temperature-dependent correlation times in the vicinity of $220\,$K, the
NMR data of all studied hydrated proteins are at variance with an existence of a sharp
FST reported in NS work for the hydration water of lysozyme at such temperatures
\cite{Chen_PNAS_06}, see Fig.\ \ref{corrtimes}. There, it was found that the
translational relaxation time of hydration water follows an Arrhenius law with a small
activation energy of $E_a=0.14\,$eV below $220\,$K, consistent with temperature-dependent
diffusivity measurements \cite{Mallamace_JCP_07}. A dynamical process with such
characteristics is not observed in our present and previous \cite{Vogel08} $^2$H NMR
studies on hydration waters. For example, if there was a FST, the low-temperature process
would be the $\alpha$ process and, hence, fast isotropic motion with $\tau\ll1/\delta$
would exist down to about $140\,$K, see Fig.\ \ref{corrtimes}. In this case, a narrow Lorentzian is expected as $^2$H NMR LS, at variance with the observations at such temperatures, see Figs.\
\ref{spectraM} and \ref{spectraE}.

Below $220\,$K, NMR and DS correlation times nicely agree, see Fig.\ \ref{corrtimes}, indicating that the same dynamical process is probed. For both techniques, the depicted correlation times describe the peak position of $G(\log \tau)$ and, hence, direct comparison should be possible. However, we have to consider that DS and NMR probe the rotational correlation functions of the first and second Legendre polynomials, respectively. The time constants of both these correlation functions coincide when the reorientation involves large jump angles, while they differ by a factor of 3 for small jump angles, i.e., for rotational diffusion \cite{Boehmer01}. Hence, the observed agreement of NMR and DS data may suggest that hydration water shows large-angle jumps below $220\,$K. For the width of the CC distributions, comparable values $\alpha_{cc}\approx0.48$ are obtained in NMR and DS \cite{Swenson_data}. Thus, both methods yield consistent distributions $G(\log \tau)$. In particular, for myoglobin, the values of $\langle T_{1w} \rangle$ measured in NMR agree with that calculated on the basis of the distributions $G(\log \tau)$ determined in DS work \cite{Swenson_data}. However, the correlation times of the present SLR analysis are about two orders of magnitude shorter than that determined in our previous $^2$H NMR stimulated-echo study \cite{Vogel08}. This apparent discrepancy can be rationalized when we consider that these NMR techniques measure different types of mean correlation times. While stimulated-echo experiments yield the mean correlation time $\langle\tau\rangle$, which is the time integral of the rotational correlation function, the peak position $\tau_{cc}$ of $G(\log \tau)$ results from the present SLR analysis. For a logarithmic Gaussian distribution of correlation times characterized by a width parameter $\sigma$, both time constants are related by \cite{Vogel02}
\begin{equation}
\log_{10}\left(\frac{\langle\tau\rangle}{\tau_{cc}}\right)=\frac{\sigma^2}{2\log_{10}\mathrm{e}}
\end{equation}
when exchange within $G(\log \tau)$ can be neglected. Using $\sigma=1.43$ from the present SLR analysis, we expect $\log_{10}\left(\langle\tau\rangle/\tau_{cc}\right)=2.35$, consistent with the experimentally observed difference of both mean values.

Above $220\,$K, DS \cite{Swenson_PRL_06,Cerveny_PRE_08,Khodadadi_JPCB_08} and NS \cite{Chen_PNAS_06,Doster_EBPJ_08} studies reported deviations from an Arrhenius law for various hydrated proteins and a Vogel-Fulcher-Tammann (VFT) equation was used to interpolate the data. In this temperature range, the present SLR analysis has suffered from lacking knowledge about the shape of the spectral density. While VFT behavior results from use of the CD spectral density, Arrhenius behavior is obtained from the CC spectral density. In any case, we have found that, although a distribution of correlation times exists, SLR is exponential above $220\,$K, indicating that the water molecules sample a representative set of local environments on the time scale of $T_{1w}$, i.e., exchange processes within $G(\log \tau)$ result in short lifetimes of the dynamical heterogeneities. Such behavior is known from $^2$H NMR SLR studies on the $\alpha$ process of supercooled liquids \cite{Boehmer01}.

Altogether, our results may suggest the following picture of the temperature-dependent dynamics of protein hydration waters. Above about $220\,$K, water exhibits isotropic rotational motion, which is governed by pronounced dynamical heterogeneities with short lifetimes. Hence, the dynamics resembles the $\alpha$ process of supercooled liquids, although the vicinity of protein surfaces may lead to some differences, e.g., in the shape of $G(\log \tau)$ \cite{Swenson_PRL_06,Cerveny_PRE_08,Khodadadi_JPCB_08}. Below about $220\,$K, water exhibits anisotropic large-amplitude rotational motion, implying existence of a water-specific $\beta$ process, which shows similar features for the hydration waters of various materials, including polymers and inorganic nanoporous matrices \cite{Cerveny_PRE_08,Swenson_PRL_04}. The crossover between both mechanisms is smooth, suggesting a direct relation of both processes. For example, the $\beta$ process may be a spatially restricted precursor of the $\alpha$ process. Alternatively, both processes may require breaking of the same number of hydrogen bonds, leading to comparable temperature dependencies and distributions $G(\log \tau)$. For example, interstitial defect diffusion cannot be excluded as origin of the low-temperature process \cite{Vogel08}. Because of the large angular displacements, the water-specific $\beta$ process destroys a large fraction of orientational correlation. Thus, only a small fraction is left for the $\alpha$ process, if existent \cite{Swenson_PRL_06}, so that it is difficult to extract information about the structural relaxation from NMR and DS studies below $220\,$K, say. Furthermore, it is an open question whether the observed change in the motional mechanism is related to an ordering of the hydrogen-bond network upon cooling, e.g., as a consequence of a proposed transition from a high-density liquid to a low-density liquid \cite{Stanley}.     

$^{13}$C CP MAS spectra have shown that hydration results in enhanced elastin dynamics at ambient temperatures. This enhancement varies among different amino acids and it is particularly prominent for Gly and Pro. Upon cooling, the enhanced elastin mobility decreases and vanishes at about $243\,$K. The finding that LS changes cease at higher temperatures in $^{13}$C NMR than in $^2$H NMR indicates that protein dynamics is slower than water dynamics. Combined application of DS and NS on hydrated lysozyme enabled identification of two relaxation processes of the protein \cite{Khodadadi_JPCB_08}. While the slower protein process exhibits much longer correlation times, the time constants of the faster process coincide with that of the water dynamics, see Fig. \ref{corrtimes}. Therefore, the faster process was ascribed to a water-coupled structural relaxation of lysozyme. The present finding that protein dynamics is slower than water dynamics implies that the faster protein process is not probed in the present $^{13}$C CP MAS spectra, suggesting that this process cannot be identified with the structural relaxation, although possible differences between elastin and lysozyme need to be studied in future work. Rather, $^{13}$C CP MAS spectra reveal a freezing of elastin dynamics faster than milliseconds at about $243\,$K. At the present, the origin of the observed protein process is an open question. Moreover, it is not yet clear whether this process involves those degrees of freedom that freeze in during the protein dynamic transition. Our finding that, at all studied temperatures, the proteins contribute a Pake spectrum to the $^2$H NMR LS shows that essentially all sub-milliseconds dynamics of deuterated protein regions have small amplitudes, in harmony with results in the literature \cite{Perry_BJ_02,Yao_MRC_04},  

\begin{acknowledgments}

We are grateful to W.\ Doster for providing us with a sample of hydrated myoglobin and for valuable discussions, to F.\ Fujara for making his unpublished NMR data on hydrated myoglobin available to us and to H. Jansson, J. Swenson, and R. Bergman for allowing us to use their unpublished DS data on hydrated myoglobin. Moreover, we appreciate financial support of the DFG through grants VO 905/3-1 and VO 905/3-2.   
\end{acknowledgments}


%
\end{document}